\documentclass{article}

\usepackage{microtype}
\usepackage{graphicx}
\usepackage{subcaption}
\usepackage{booktabs} 
\usepackage{enumitem}

\usepackage{hyperref}

\usepackage[accepted]{icml2026}

\usepackage{amsmath}
\usepackage{amssymb}
\usepackage{mathtools}
\usepackage{amsthm}

\theoremstyle{definition}

\usepackage[capitalize,noabbrev]{cleveref}

\theoremstyle{plain}

\theoremstyle{definition}

\theoremstyle{remark}

\usepackage[textsize=tiny]{todonotes}

\begin{document}

\twocolumn[
  \icmltitle{Human–AI Co-Evolution and Epistemic Collapse: A Dynamical Systems Perspective}



  \icmlsetsymbol{equal}{*}

\begin{icmlauthorlist}
    \icmlauthor{Xuening Wu}{yyy,comp}
    \icmlauthor{Yanlan Kang}{yyy}
    \icmlauthor{Qianya Xu}{ucsd}
    \icmlauthor{Kexuan Xie}{nju}
    \icmlauthor{Jiaqi Mi}{hku}
    \icmlauthor{Honggang Wang}{tong}
    \icmlauthor{Yubin Liu}{jia}
    \icmlauthor{Zeping Chen}{tong}
\end{icmlauthorlist}

\icmlaffiliation{yyy}{Fudan University, Shanghai, China}
\icmlaffiliation{comp}{Pfizer China, Shanghai, China}
\icmlaffiliation{tong}{Tongji University, Shanghai, China}
\icmlaffiliation{jia}{Shanghai Jiao Tong University, Shanghai, China}
\icmlaffiliation{hku}{The University of Hong Kong, Hong Kong, China}
\icmlaffiliation{ucsd}{University of California San Diego, CA, USA}
\icmlaffiliation{nju}{Nanjing University of Posts and Telecommunications, Nanjing, China}

\icmlcorrespondingauthor{Xuening Wu}{xueningwu@gmail.com}

  \icmlkeywords{Epistemic Intelligence, Human-AI Co-evolution, Model Collapse, 
Epistemic Uncertainty, Dynamical Systems, AI Safety}

  \vskip 0.3in
]

\printAffiliationsAndNotice{}  

\begin{abstract}

Large language models (LLMs) are reshaping how knowledge is produced, with increasing reliance on AI systems for generation, summarization, and reasoning. While prior work has studied cognitive offloading in humans and model collapse in recursive training, these effects are typically considered in isolation.
We propose a unified perspective: humans and language models form a coupled dynamical system linked by a feedback loop of usage, generation, and retraining. We introduce a minimal model with three variables—human cognition, data quality, and model capability—and show that this feedback can give rise to distinct dynamical regimes.
Our analysis identifies three regimes: co-evolutionary enhancement, fragile equilibrium, and degenerative convergence. Through a simple simulation, we demonstrate that increasing reliance on AI can induce a transition toward a low-diversity, suboptimal equilibrium.
From an information-theoretic perspective, this transition corresponds to an emergent information bottleneck in the human–AI loop, where entropy reduction reflects loss of diversity and support under closed-loop feedback rather than beneficial compression. These results suggest that the trajectory of AI systems is shaped not only by model design, but by the dynamics of human–AI co-evolution.
\end{abstract}


\section{Introduction}

The rise of large language models (LLMs) is reshaping how knowledge is produced. Tasks that once required sustained human reasoning—such as writing, summarization, and synthesis—can now be partially or fully outsourced to AI systems. While this shift improves efficiency, it raises a fundamental question: what happens when humans and models adapt to each other over time?

Two lines of evidence motivate this question. On the human side, reliance on AI tools reduces cognitive engagement and memory formation \cite{kosmyna2025brain, sparrow2011google}. On the model side, training on recursively generated data leads to distributional degeneration and loss of diversity \citep{shumailov2024model}. While these effects are typically studied in isolation, we argue that they instead reflect a single coupled process \cite{rahwan2019machine}.

We study \emph{epistemic intelligence}—the capacity of human–AI systems to generate and sustain high-quality knowledge—as a dynamical property of a feedback system. We propose a minimal model in which human cognition and language models are coupled: humans increasingly rely on models for cognitive tasks, while models are trained on human--AI-generated data.
This feedback loop gives rise to qualitatively distinct regimes, including enhancement, equilibrium, and degeneration. 

This framework admits an information-theoretic interpretation in which degeneration corresponds to an emergent information bottleneck\cite{tishby2000information}, i.e., a progressive restriction of the effective support of the data distribution. Unlike beneficial compression in open systems, the associated entropy reduction in the closed human–AI loop reflects contraction of the data distribution and loss of novel information.

We introduce a minimal formal framework that captures this feedback structure, exhibits transitions between regimes, and provides a basis for analyzing long-term human–AI co-evolution. This perspective reframes model collapse as a coupled dynamical phenomenon, rather than a purely data-centric issue.

\begin{figure*}[t]
    \centering
    \includegraphics[width=0.75\linewidth]{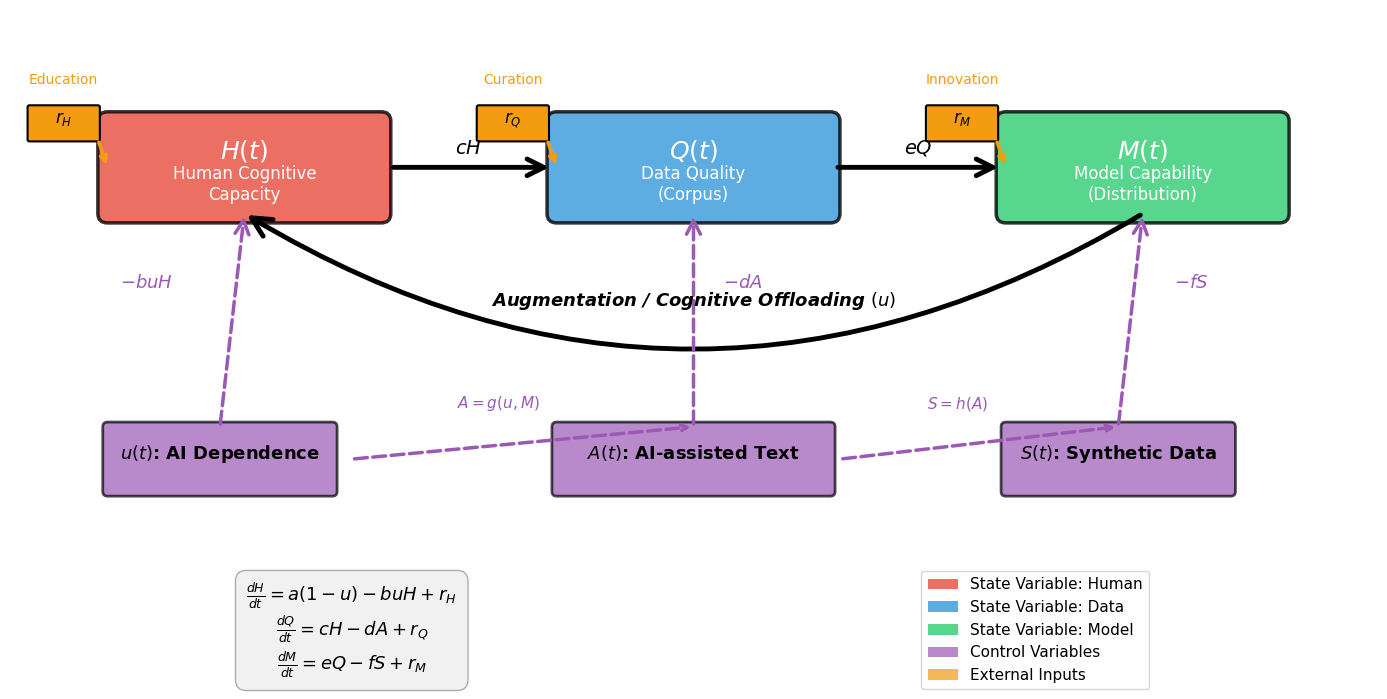}
    \caption{\textbf{The closed-loop dynamics of epistemic intelligence.} 
The diagram illustrates the coupled ODE system, with state variables 
$H(t), Q(t), M(t)$ (top) linked by forward enhancement pathways (solid arrows) 
and control variables $u, A, S$ (bottom) introducing negative feedback via 
cognitive offloading and recursive data contamination (dashed arrows). 
External inputs $r_H, r_Q, r_M$ represent intervention mechanisms. The feedback 
structure $H \to Q \to M \to H$ captures the co-evolution of human cognition, data quality, and model capability.}
    \label{fig:close_loop}
\end{figure*}

\section{A Coupled Dynamical Model}
We define the system state as $\mathbf{x}(t) = [H, Q, M]^\top \in \mathbb{R}^3_+$, representing human cognitive capacity ($H$), collective data quality ($Q$), and model capability ($M$). These variables are coupled via a feedback loop $H \to Q \to M \to H$ governed by the following ODE system:
\begin{equation}
\label{eq:dynamics}
\begin{cases}
    \dot{H} = a(1-u) - b\,uH + r_H \\
    \dot{Q} = c H - d A + r_Q \\
    \dot{M} = e Q - f S + r_M
\end{cases}
\end{equation}
where $u \in [0,1]$ is the degree of cognitive offloading. The system is closed by control relations $A = g(u,M)$ and $S = h(A)$, instantiated in \cref{sec:simulation} as $A = \alpha u M$ and $S = \beta A$.

\noindent \textbf{Parameters:} (i) \textbf{Reinforcement:} $a, c, e > 0$ capture cognitive growth, human-driven quality, and model scaling; (ii) \textbf{Degradation:} $b, d, f > 0$ model offloading decay, content drift, and synthetic noise; (iii) \textbf{Exogenous:} $r_H, r_Q, r_M$ represent interventions such as education, curation, and innovation. This formulation captures the risk of a "divergence shift," where increasing AI dependence amplifies recursive data generation, potentially decoupling model performance from real-world utility.

\subsection{Fixed Points and Stability}

A fixed point $\mathbf{x}^* = (H^*, Q^*, M^*)$ satisfies $\dot{H} = \dot{Q} = \dot{M} = 0$. Solving yields
\begin{align}
    H^* &= \frac{a(1-u) + r_H}{bu}, \\
    Q^* &= \frac{1}{e} \left( f \beta \alpha u M^* - r_M \right),
\end{align}
where $M^*$ is determined by the remaining equation. Local stability is governed by the eigenvalues of the Jacobian $\mathbf{J} = \nabla \mathbf{f}(\mathbf{x}^*)$.

As the offloading parameter $u$ increases, the system exhibits a \textit{transcritical bifurcation} at a critical threshold $u_c$. For $u < u_c$, a stable high-capability equilibrium yields a \textbf{Co-evolutionary Enhancement} regime. For $u > u_c$, this equilibrium loses stability, and the system converges to a \textbf{Degenerative Convergence} attractor with reduced epistemic diversity.

\subsection{Three Regimes}
\label{sec:regimes}

We identify three qualitatively distinct regimes, arising from different balances between reinforcing and degrading feedback in the coupled system.

\begin{figure*}[t]
    \centering
    \includegraphics[width=1\linewidth]{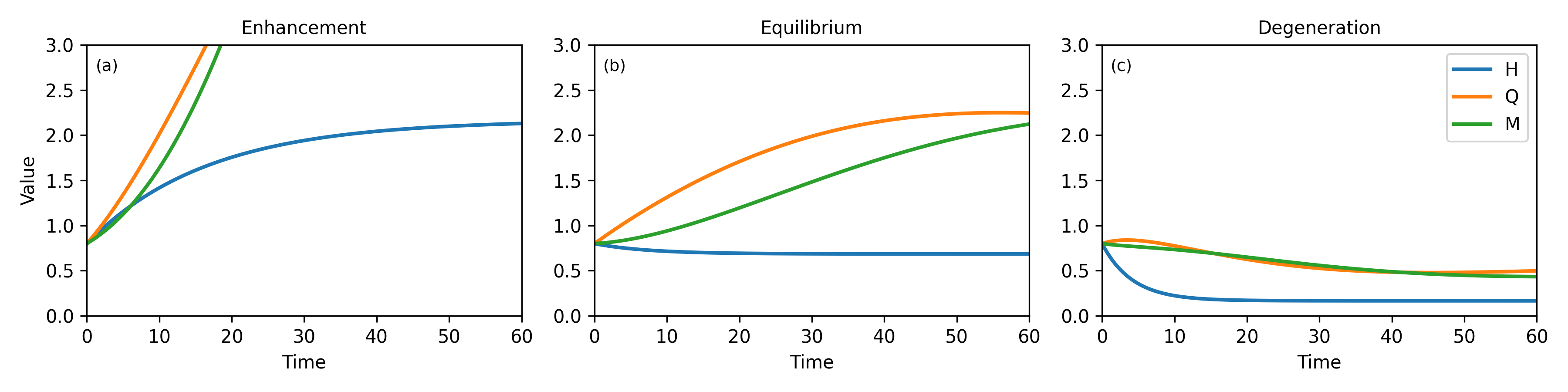}
\caption{
Simulation of the coupled human–data–model system under three representative regimes.
(a) \textbf{Enhancement}: all variables increase over time due to dominant positive feedback among human cognition ($H$), data quality ($Q$), and model capability ($M$).
(b) \textbf{Equilibrium}: the system converges to a bounded fixed point, reflecting a balance between reinforcing and degrading feedback.
(c) \textbf{Degeneration}: human cognition and data quality decline, while model capability stabilizes at a low-diversity equilibrium due to increasing reliance on synthetic data.
}
    \label{fig:regimes}
\end{figure*}

\subsubsection{Co-evolutionary Enhancement}

When cognitive engagement remains high (low $u$) and data quality is preserved (low $A, S$), the system exhibits sustained growth:
\[
H \uparrow, \quad Q \uparrow, \quad M \uparrow.
\]
In this regime, positive feedback dominates, and the system does not converge to a finite fixed point within the relevant time horizon.

\subsubsection{Fragile Equilibrium}

Under moderate AI dependence and active external correction (e.g., curation or training), the system converges to a bounded fixed point:
\[
(H^*, Q^*, M^*).
\]
This corresponds to a locally stable equilibrium, defining a metastable regime in which performance remains acceptable but long-term growth is limited. Intuitively, this corresponds to a parameter regime in which the local Jacobian at $(H^*, Q^*, M^*)$ has eigenvalues with negative real parts, so that perturbations decay over time.

\subsubsection{Degenerative Convergence}

When $u$, $A$, and $S$ exceed critical levels and corrective inputs $r_H, r_Q, r_M$ are insufficient, the system enters a degenerative regime:
\[
H \downarrow, \quad Q \downarrow, \quad M \rightarrow M^*,
\]
where $M^*$ corresponds to a low-diversity equilibrium.

In this regime, degrading feedback dominates: reduced human contribution and recursive data contamination drive the system toward a stable but low-diversity attractor.

\subsection{Information-Theoretic Interpretation}

The dynamics admit an information-theoretic interpretation. As cognitive offloading increases, the entropy of human-generated content decreases, reducing diversity in the shared corpus. This induces an information bottleneck along the feedback loop $H \rightarrow Q \rightarrow M \rightarrow H$, limiting the system’s capacity to incorporate novel information.

In the degenerative regime, this manifests as a divergence shift, where the model distribution aligns more closely with human-generated data while drifting away from the underlying world distribution, resulting in progressive information loss and reduced sensitivity to external knowledge.

\section{Testable Predictions}

The proposed dynamical framework yields the following empirically testable predictions:

\textbf{(1) Human cognition.}
Increasing AI dependence ($u$) leads to reduced entropy and lexical diversity in human-generated text, reflecting diminished cognitive engagement.

\textbf{(2) Data distribution.}
As the proportion of synthetic content increases, the effective data distribution exhibits entropy reduction and contraction of the long tail, indicating loss of rare or novel patterns.

\textbf{(3) Model performance.}
Models trained on recursively generated data exhibit degradation in out-of-distribution generalization, despite maintaining or improving in-distribution performance.

\textbf{(4) Regime transition.}
There exists a critical threshold in AI dependence ($u_c$) beyond which the system transitions from a co-evolutionary enhancement regime to a degenerative regime, observable as a sharp change in steady-state metrics (e.g., $Q^*$ or $M^*$).

\section{Simulation of the Coupled Dynamics}
\label{sec:simulation}

To complement the theoretical analysis, we simulate the proposed coupled dynamical system to illustrate the emergence of the regimes described in Section~\ref{sec:regimes}.

We adopt linear control relations of the form $A = \alpha u M$ and $S = \beta A$, where $u \in [0,1]$ denotes the degree of cognitive offloading. The system is integrated using forward Euler discretization with a fixed step size.

Our objective is not precise quantitative modeling, but to demonstrate that the feedback structure alone is sufficient to generate qualitatively distinct behaviors.

We consider three representative parameter settings corresponding to enhancement, equilibrium, and degeneration. The resulting trajectories of $H(t)$, $Q(t)$, and $M(t)$ are shown in \cref{fig:regimes}. 

The simulations exhibit three characteristic regimes: (i) joint growth driven by reinforcing feedback, (ii) convergence to a bounded fixed point under balanced feedback, and (iii) collapse toward a low-diversity attractor under dominant degrading feedback. These behaviors are consistent with the qualitative predictions of the model.

\section{Discussion}

Our analysis suggests that the primary risk of AI is not merely capability misalignment, but \emph{dynamical misalignment}—a degradation of epistemic intelligence arising from coupled human–model adaptation. In this view, failure is systemic rather than component-wise, emerging from feedback loops rather than isolated model behavior.

In the degenerative regime, the system approaches a state of epistemic closure: $M^*$ reflects not only reduced diversity, but also a diminished capacity to incorporate novel external information, effectively collapsing sensitivity to out-of-distribution knowledge. The system becomes increasingly self-referential, reinforcing its own outputs rather than tracking the underlying world distribution.

Thus, the central issue is not model capability per se, but the dynamics induced by feedback between human cognition, data generation, and model training.

\begin{figure}[t]
    \centering
    \includegraphics[width=1\linewidth]{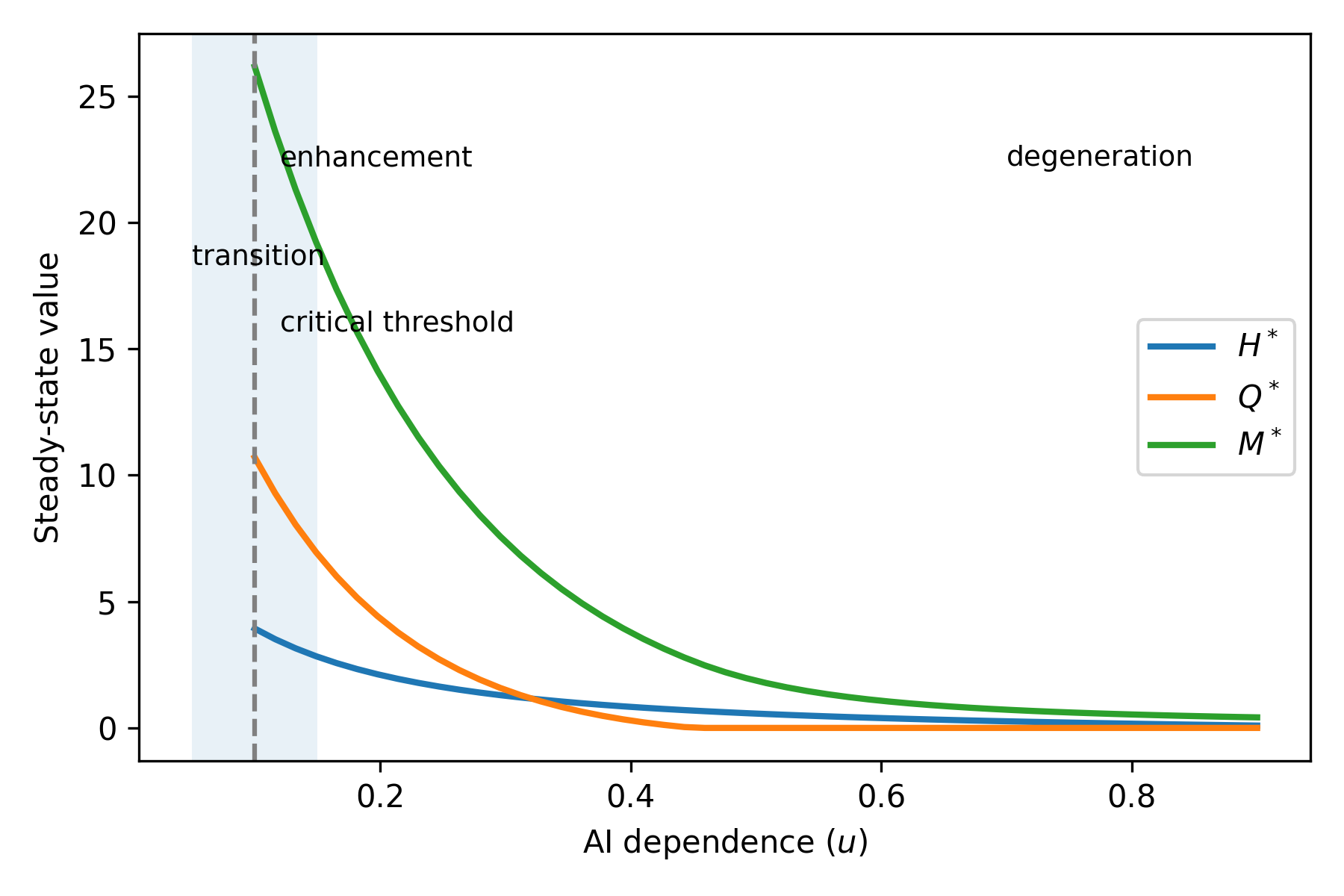}
    \caption{
Steady-state behavior of the coupled system as a function of AI dependence $u$. 
As $u$ increases, the system transitions from an enhancement regime to a degenerative equilibrium. 
The dashed line indicates the critical threshold $u_c$, identified by the maximal gradient change in $Q^*(u)$. 
The shaded region denotes the transition band separating distinct dynamical regimes. 
The monotonic decline of $H^*$, $Q^*$, and $M^*$ reflects the increasing dominance of degrading feedback under higher AI dependence.
}
    \label{fig:threshold}
\end{figure}

\subsection{Information-Theoretic Perspective}

Model collapse can be understood as a coupled socio-technical phenomenon rather than a purely data-driven issue. 
From an information-theoretic perspective, the degenerative regime corresponds to a divergence shift:
\[
D_{\mathrm{KL}}(P_{\mathrm{model}} \,\|\, P_{\mathrm{human}}) \downarrow
\quad \text{while} \quad
D_{\mathrm{KL}}(P_{\mathrm{model}} \,\|\, P_{\mathrm{world}}) \uparrow.
\]
This reflects increasing alignment with human-generated distributions, coupled with drift away from the true data-generating process.

Recursive contamination induces an \emph{echo chamber effect}, in which the effective data distribution collapses toward a low-entropy attractor. As a result, information transmission across generations degrades:
\[
I(H_t; H_{t+1}) < I(H_{t-1}; H_t).
\]

\subsection{Intervention Strategies}

The model suggests three intervention points corresponding to the state variables of the system:
\begin{itemize}
    \item Increasing $r_H$ (human cognition): through education and practices that preserve active reasoning and cognitive engagement.
    \item Increasing $r_Q$ (data quality): via curation, filtering, and incorporation of high-quality human or expert-generated data.
    \item Increasing $r_M$ (model capability): through architectural improvements that enhance robustness to distributional shift and synthetic data contamination.
\end{itemize}

These results suggest that alignment must be understood not only as a property of models, but as a property of the coupled human–AI system.

\subsection{Limitations.}
The present work is intentionally minimal and conceptual. 
The proposed ODE system is not intended as a quantitatively validated model of human cognition or societal dynamics, but rather as a formal framework for studying possible feedback mechanisms in human--AI co-evolution. 
Future work will empirically investigate entropy reduction, recursive semantic collapse, and diversity contraction under iterative AI-mediated generation.

\section{Conclusion}

We introduce a minimal dynamical framework for analyzing long-term human–AI co-evolution, unifying cognitive offloading and model collapse within a single feedback system. Our analysis shows that the same feedback structure can give rise to enhancement, equilibrium, or degeneration, depending on the level of AI dependence.

This perspective suggests that the trajectory of AI systems is governed not only by model design, but by the coupled dynamics of human cognition, data generation, and model training.

More broadly, alignment should be understood as a property of the human–AI system as a whole, rather than of models in isolation.

\section*{Impact Statement}

This work presents a conceptual framework for analyzing the long-term co-evolution of human cognition and generative models as a coupled dynamical system. A central implication is that increasing reliance on AI, combined with recursive data generation, may degrade both human cognitive engagement and data quality, potentially driving the system toward low-diversity, suboptimal equilibria.

If such dynamics occur in practice, they could contribute to reduced creativity, diminished critical thinking, and homogenization of knowledge production, particularly in high-stakes domains such as scientific research, education, and decision-making.

At the same time, the framework highlights potential mitigation strategies, including improved data curation, human-in-the-loop system design, and educational practices that preserve active cognitive engagement. These interventions can be interpreted as control mechanisms that steer the system toward a co-evolutionary enhancement regime.

This work is primarily theoretical and supported by minimal simulations. Its main contribution is to provide testable hypotheses and a systems-level perspective for designing human–AI systems that preserve epistemic diversity and avoid degeneration under feedback dynamics.

\bibliography{example_paper}
\bibliographystyle{icml2026}

\end{document}